\documentclass[mathleft]{an}
\usepackage{graphicx,times}
\usepackage{bm}
\graphicspath{{./fig/}{./png/}}
\Yearpublication{2012}
\Yearsubmission{2011}
\Month{7}
\Volume{333}
\Issue{2}
\DOI{10.1002/asna.201111638}
\Pagespan{95}{}
\newcommand{\EQ}{\begin{equation}}
\newcommand{\EN}{\end{equation}}
\newcommand{\EQA}{\begin{eqnarray}}
\newcommand{\ENA}{\end{eqnarray}}
\newcommand{\eq}[1]{(\ref{#1})}

\newcommand{\Eq}[1]{Eq.~(\ref{#1})}

\newcommand{\Fig}[1]{Fig.~\ref{#1}}

\newcommand{\Tab}[1]{Table~\ref{#1}}

\newcommand{\meanrho}{\overline{\rho}}

{}
{}
\newcommand{\meanFFFF}{\overline{\mbox{\boldmath ${\cal F}$}}{}}{}

\newcommand{\meanSSSS}{\overline{\mbox{\boldmath ${\mathsf S}$}} {}}

{}
{}
{}
{}
{}
{}
{}
\newcommand{\meanAA}{\overline{\mbox{\boldmath $A$}}{}}{}
\newcommand{\meanBB}{\overline{\mbox{\boldmath $B$}}{}}{}
{}
{}
{}
{}
{}
{}
{}
\newcommand{\meanJJ}{\overline{\mbox{\boldmath $J$}}{}}{}
\newcommand{\meanUU}{\overline{\bm{U}}}

{}
{}
{}
\newcommand{\meanA}{\overline{A}}
\newcommand{\meanB}{\overline{B}}

\newcommand{\meanU}{\overline{U}}

\newcommand{\meanp}{\overline{p}}

{}

{}
{}

\newcommand{\uu}{\mbox{\boldmath $u$} {}}
\newcommand{\UU}{\mbox{\boldmath $U$} {}}

\newcommand{\BB}{\mbox{\boldmath $B$} {}}

\newcommand{\grav}{\mbox{\boldmath $g$} {}}
\newcommand{\nab}{\mbox{\boldmath $\nabla$} {}}
\newcommand{\const}{{\rm const}  {}}

\def\cs{c_{\rm s}}
\def\kd{k_{\rm d}}

\def\qs{q_{\rm s}}
\def\qsz{q_{\rm s0}}
\def\qpz{q_{\rm p0}}
\def\betap{\beta_{\rm p}}
\def\betacrit{\beta_{\rm crit}}

\def\qp{q_{\rm p}}
\def\qpz{q_{\rm p0}}

\def\kf{k_{\rm f}}

\def\urms{u_{\rm rms}}

\def\nut{\nu_{\rm t}}
\def\etat{\eta_{\rm t}}

\def\Beq{B_{\rm eq}}
\def\Beqz{{B_{\rm eq0}}}
\def\betap{\beta_{\rm p}}
\def\betamin{\beta_{\min}}
\def\betastar{\beta_{\star}}
\def\betastarz{\beta_{\star0}}
\def\Peff{{\cal P}_{\rm eff}}
\def\Pmin{{\cal P}_{\rm min}}

\def\half{{\textstyle{1\over2}}}

\def\onethird{{\textstyle{1\over3}}}

\newcommand{\yapj}[3]{: #1, {ApJ} {#2}, #3}
\newcommand{\yapjl}[3]{: #1, {ApJ} {#2}, #3}

\newcommand{\yan}[3]{: #1, {AN} {#2}, #3}

\newcommand{\yana}[3]{: #1, {A\&A} {#2}, #3}

\newcommand{\ygafd}[3]{: #1, {Geophys. Astrophys. Fluid Dyn.} {#2}, #3}

\newcommand{\ypf}[3]{: #1, {Phys. Fluids} {#2}, #3}
\newcommand{\ypfb}[3]{ #1, {Phys.\ Fluids B,} {#2}, #3}

\newcommand{\ysovl}[3]{: #1, {Sov. Astron. Lett.} {#2}, #3}
\newcommand{\yjetp}[3]{: #1, {Sov. Phys. JETP} {#2}, #3}
\newcommand{\yphy}[3]{: #1, {Physica} {#2}, #3}

\newcommand{\yprs}[3]{: #1, {Proc. Roy. Soc. Lond.} {#2}, #3}

\newcommand{\ypre}[3]{: #1, {Phys. Rev. E} {#2}, #3}

\newcommand{\ssph}[1]{: #1, {Sol.\ Phys.}, submitted}
\newcommand{\sapj}[1]{: #1, {ApJ}, submitted}

\newcommand{\smn}[1]{: #1, {MNRAS, submitted}}

\title{Properties of the negative effective magnetic pressure instability}
\titlerunning{Negative effective magnetic pressure instability}
\authorrunning{K. Kemel et al.}
\author{K.\ Kemel\inst{1,2}\fnmsep\thanks{Corresponding author:
 koentjekemel@hotmail.com}\and A.\ Brandenburg\inst{1,2}\and N.\
 Kleeorin\inst{3,1}\and I.\ Rogachevskii\inst{3,1}}
\institute{
Nordita\thanks{Nordita is a Nordic research institute jointly operated
by the Stockholm University and the Royal Institute of Technology, Stockholm.},
AlbaNova University Center, Roslagstullsbacken 23,
SE 10691 Stockholm, Sweden
\and
Department of Astronomy, AlbaNova University Center,
Stockholm University, SE 10691 Stockholm, Sweden
\and
Department of Mechanical
Engineering, The Ben-Gurion University of the Negev, POB 653,
Beer-Sheva 84105, Israel
}
\received{2011 Jul 14} \accepted{2011 Nov 28} \publonline{2012 Feb 15}

\keywords{magnetohydrodynamics (MHD) -- instabilities -- turbulence}
\abstract{%
As was demonstrated in earlier studies, turbulence can result in a
negative contribution to the effective mean magnetic pressure,
which, in turn, can cause a large-scale instability.
In this study, hydromagnetic mean-field modelling
is performed
for an isothermally stratified layer in the presence
of a horizontal magnetic field.
The negative effective magnetic pressure
instability (NEMPI) is comprehensively investigated.
It is shown that, if the effect of turbulence on the mean
magnetic tension force vanishes,
which is consistent with results from direct numerical
simulations of forced turbulence, the fastest growing
eigenmodes of NEMPI are two-dimensional.
The growth rate is found to depend on a parameter $\betastar$ characterizing
the turbulent contribution of the effective mean magnetic pressure for
moderately strong mean magnetic fields.
A fit formula is proposed that gives
the growth rate as a function of turbulent kinematic
viscosity, turbulent magnetic diffusivity,
the density scale height, and the parameter $\betastar$.
The strength of the imposed magnetic field does not explicitly enter
provided the location of the vertical boundaries are chosen such
that the maximum of the eigenmode of NEMPI fits into the domain.
The formation of sunspots and solar active regions
is discussed as possible applications of NEMPI.
}

\begin{document}
\maketitle

\section{Introduction}

The concept of turbulent viscosity is often used in astrophysical and other
applications in recognition of the fact that the microscopic viscosity is
far too small to be relevant on the length scales under consideration.
Turbulent viscosity is the simplest parameterization of the Reynolds
stress tensor, $\overline{u_i u_j}$, where $\uu=\UU-\meanUU$ is the
velocity fluctuation about a suitably defined average, denoted here by
an overbar.
Turbulent viscosity is by far not the only contribution to the Reynolds stress tensor.
In addition to hydrodynamic contributions such as the $\Lambda$ effect
(R\"udiger 1980, 1989), which is relevant to explaining stellar differential
rotation (R\"udiger \& Hollerbach 2004),
and the anisotropic kinetic alpha effect (Frisch et al.\ 1987),
which provides an important test case in mean-field hydrodynamics
(Brandenburg \& von Rekowski 2001; Courvoisier et al.\ 2010), there are
magnetic contributions as well.
One can think of them as a magnetic feedback on the hydrodynamic stress
tensor (R\"adler 1974; R\"udiger 1974) or, especially when magnetic
fluctuations are also considered, as a mean-field contribution to the
turbulent Lorentz force.

Work by Roberts \& Soward (1975) and
R\"{u}diger et al.\ (1986, 2012) using a quasi-linear approach
suggests that the total magnetic tension force (which includes the effects
of fluctuations) is reduced in the presence of a mean magnetic field
and might formally even change sign for larger magnetic Reynolds numbers,
but this would then be beyond the validity of their approximation.
Using the spectral $\tau$ approach, Kleeorin et al.\ (1989, 1990)
do indeed find this reversal of the sign of the total magnetic tension force.
In addition, they find a reversal of the sign of the effective
magnetic {\it pressure} term;
see also Kleeorin \& Rogachevskii (1994) and Kleeorin et al.\ (1993, 1996).
Rogachevskii \& Kleeorin (2007) argue that, in a stratified medium,
this can lead to the formation of
large-scale magnetic flux structures and perhaps even sunspots --
or at least active regions.

Recently, direct numerical simulations (DNS) of both unstratified
and stratified forced turbulence (Brandenburg et al.\ 2010, 2012;
hereafter referred to as BKR and BKKR, respectively) have
substantiated this idea and have demonstrated that the effective magnetic
pressure can indeed change sign.
Similar results have now also been obtained for turbulent convection
(K\"apyl\"a et al.\ 2012).
In addition, these papers give results of mean-field calculations illustrating that
there is a negative effective magnetic pressure instability (hereafter
referred to as NEMPI) when there is sufficient density stratification.

NEMPI is a convective type instability related to the interchange
instability in plasmas (Tserkovnikov 1960; Newcomb 1961; Priest 1982) and
the magnetic buoyancy instability in the astrophysical context (Parker 1966).
The free energy in interchange and magnetic buoyancy instabilities is
drawn from the gravitational field, while in NEMPI it is provided by
the small-scale turbulence.

The mechanism of NEMPI works even under isothermal conditions
when entropy evolution is ignored and an isothermal
equation of state is used.
This has been shown using corresponding mean-field calculations (BKKR).
With this reduction to the most elementary aspects of the instability,
it has recently been possible to verify the existence of NEMPI also in DNS
(Brandenburg et al.\ 2011, hereafter referred to as BKKMR).
This has been a major step forward, because now there is
no doubt that one is pursuing a real effect and not just
one that works only in the world
of mean-field models.
Essential to the paper of BKKMR
has been a finding from an earlier version of the present one that only
two-dimensional mean-field structures are excited.
This property allowed meaningful averaging along the direction of the
imposed field, making the identification of flux concentrations thus
much clearer.
The absence of three-dimensional mean-field structures
was surprising because three-dimensional mean-field calculations
have shown that the mean magnetic field
develops structures along the direction of the imposed field (BKR).
However, while the mean-field calculations have
illustrated the nature of
the instability, no systematic survey of solutions
has yet been attempted.
The purpose of this paper is therefore to clarify some still
puzzling aspects concerning NEMPI.
Note also that the large-scale flux concentrations
observed in DNS of BKKMR have an amplitude of only 15\,\%
of the local equipartition field. This implies that
the flux concentrations we observe in DNS
are often not strong enough to be noticeable without averaging.

In addition to the structures found in BKKMR, other
types of structures have recently been
reported in Large-Eddy Simulations (LES),
which might also be an indication of NEMPI.
We have here in mind the radiation magneto-convection
simulations of Kitiashvili et al.\ (2010), in which one sees
the formation of whirlpool-like magnetic structures.
Relevant to NEMPI is also the work of Tao et al.\ (1998),
who considered magneto-convection in the optically thick approximation
and find a horizontal segregation into magnetized and non-magnetized regions.
The size of the individual regions is such that they encompass several
turbulent eddies.
This phenomenon might therefore well be associated with an effect that
could also be modelled in terms of mean-field theory.
However, before we can make such an association,
we need to find out more about the properties of NEMPI.
In particular, we need to know what is the optimal magnetic field strength,
what are the requirements or restrictions on the turbulent velocity,
and, finally, how much density stratification is needed to make NEMPI work.

To connect the aforementioned requirements to DNS, we need to have
a meaningful parameterization of the turbulence effects.
The work done so far has been focussing on measuring a reduction
of the turbulent pressure and effective mean magnetic
pressure as a function of the local mean magnetic field strength.
The shape of the resulting dependence of the effective mean magnetic
pressure on the mean magnetic field has been matched to a specific fit
formula that can be characterized by two fit parameters that, in turn,
can be linked to the minimum effective mean magnetic
pressure and critical field strength above which the effect is suppressed.
However, there have been indications that this parameterization
is not unique and that different combinations of the two fit parameters
can result in similar values of minimum effective pressure and the
critical field strength.
The question therefore arises whether this apparent degeneracy affects
the properties of NEMPI.

We mentioned already the fact that NEMPI is capable of
exciting three-dimensional structures that show variation along the
direction of the mean magnetic field.
This would give rise to the worry that the two-dimensional results presented
so far may not reflect the properties of the fastest growing mode and
may therefore not be relevant to describing NEMPI.
However, as will be discussed in this paper,
this is not the case, because the degree
to which three-dimensional modes are excited depends on the sign
and magnitude of one
of the turbulence parameters, namely the term characterizing turbulence
effects on the magnetic tension force, and that simulations indicate
that this sign is not favorable for exciting three-dimensional modes
(BKKR, K\"apyl\"a et al.\ 2012).
Before we begin addressing the various points, we discuss first the
mean-field model and basic setup.

\section{Mean-field model}

In view of further verifications of NEMPI with DNS, it is necessary
to be able to reduce the essential physics to a minimum.
We will therefore not make any attempt to consider other aspects
that would make the model more realistic with respect to the Sun.
Given that NEMPI works even under isothermal conditions (BKKR), we
adopt an isothermal equation of state where the mean pressure $\meanp$ is
linear in the mean density $\meanrho$, with $\meanp=\meanrho\cs^2$
and $\cs$ being the constant isothermal sound speed.
We solve the evolution equations for mean velocity $\meanUU$,
mean density $\meanrho$, and mean vector potential $\meanAA$, in the form
\EQ
{\partial\meanUU\over\partial t}=-\meanUU\cdot\nab\meanUU
-\cs^2\nab\ln\meanrho+\grav+\meanFFFF_{\rm M}+\meanFFFF_{\rm K},
\EN
\EQ
{\partial\meanrho\over\partial t}=-\meanUU\cdot\nab\meanrho
-\meanrho\nab\cdot\meanUU,
\EN
\EQ
{\partial\meanAA\over\partial t}=\meanUU\times\meanBB-(\etat+\eta)\meanJJ,
\EN
where $\meanFFFF_{\rm M}$ is given by
\EQ
\meanrho \, \meanFFFF_{\rm M} = -\half\nab[(1-q_{\rm p})\meanBB^2]
+ \meanBB \cdot \nab\left[(1-q_{\rm s})\meanBB\right]\!,
\label{efforce}
\EN
and
\EQ
\meanFFFF_{\rm K}=(\nut+\nu)\left(\nabla^2\meanUU+\onethird\nab\nab\cdot\meanUU
+2\meanSSSS\nab\ln\meanrho\right)
\EN
is the total (turbulent plus microscopic) viscous force.
Here, ${\sf S}_{ij}=\half(\meanU_{i,j}+\meanU_{j,i})
-\onethird\delta_{ij}\nab\cdot\meanUU$
is the traceless rate of strain tensor of the mean flow.
As in earlier work (BKR, BKKR), we approximate $\qp$ and $\qs$
by simple profiles that are only functions of the ratio
$\beta\equiv|\meanBB|/\Beq$.
However, in the earlier work this functional form was described by
\EQ
q_\sigma(\beta)=q_{\sigma 0}[1-(2/\pi)
\arctan(\beta^2/\beta_\sigma^2)],
\label{qp}
\label{qs}
\EN
where $\Beq$ is the equipartition field strengths and
$\sigma$ stands for subscripts p and s, respectively.
We refer to this as the arctan fit.
In the present paper we use an algebraic fit of the form
\EQ
q_\sigma(\beta)=q_{\sigma 0}[1-(2/\pi)
\arctan(\beta^2/\beta_\sigma^2)],
\label{qp}
\label{qs}
\EN
\begin{equation}
q_\sigma(\beta)={q_{\sigma 0}\over1+\beta^2/\beta_\sigma^2}.
\label{qpbeta2}
\end{equation}
The functions $\qp$ and $\qs$ quantify the impact of the mean
magnetic field
on the effective pressure and tension forces, respectively.

\begin{figure*}[t!]\begin{center}
\includegraphics[width=.9\textwidth]{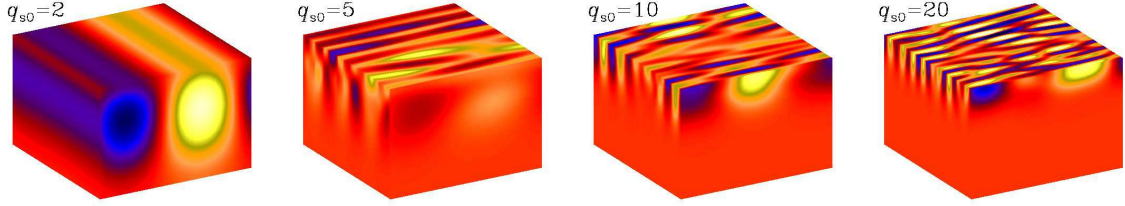}
\end{center}\caption[]{
(online colour at: www.an-journal.org) Visualization of $\meanB_y$ at the periphery of the computational domain
near the end of the kinematic growth phase.
Note the change of the field pattern with increasing values of
$\qsz =\ $2, 5, 10, and 20 (\emph{from left to right}).
}\label{b003_ly2pi_qs0}\end{figure*}

As initial condition, we assume a hydrostatic stratification with
${\meanrho(z)=\rho_0\exp(-z/H_\rho)}$, where
${H_\rho=\cs^2/g}$ is the scale height
in our domain of size $L_x{\times} L_y{\times} L_z$; the exact dimensions
vary between 4 and 10 density scale heights in each direction.
We add a small perturbation to the velocity field.
We allow for the presence of an imposed field
in the $y$ direction, $\BB_0=(0,B_0,0)$.
The total field is then written as
\EQ
\meanBB=\BB_0+\nab\times\meanAA,
\EN
so the departure from the imposed field is expressed in terms
of the mean magnetic vector potential $\meanAA$.
Furthermore, we assume
\EQ
\Beq(z)=\Beqz\exp(-z/2H_\rho),
\EN
with a normalization coefficient $\Beqz$.
This formula is compatible with $\Beq=\rho^{1/2}\urms$ in BKKMR,
where the turbulent rms velocity, $\urms$, was approximately constant.

On the upper and lower boundaries
we adopt stress-free boundary conditions for velocity,
i.e.\ $\meanU_{x,z}=\meanU_{y,z}=\meanU_{z}=0$, and a perfect conductor
boundary condition for the magnetic field, i.e.\
${\meanA_x=\meanA_y=\meanA_{z,z}=0}$.
Here, commas denote partial differentiation.
No boundary condition for the density is required.
All computations have been carried out with the {\sc Pencil Code}\footnote{
http://www.pencil-code.googlecode.com}.

Our model is characterized by the following set of input parameters.
There are three parameters characterizing the hydrostatic equilibrium
stratification, namely $g$, $\cs$ and $\rho_0$.
The remaining parameters are the normalized imposed field strength,
$B_0/\Beqz$, turbulent viscosity and magnetic
diffusivity, as well as the parameters $q_{\sigma0}$ and $\beta_{\sigma}$.

\section{Results}

\subsection{Two- and three-dimensional solutions}

\begin{figure}[t!]\begin{center}
\includegraphics[width=.9\columnwidth]{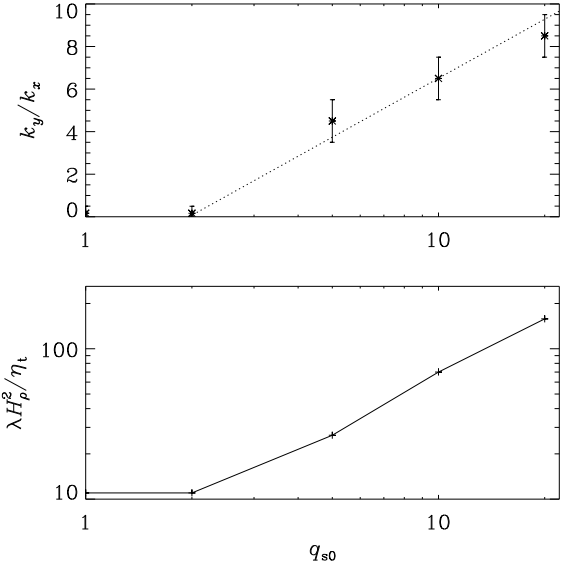}
\end{center}\caption[]{
Dependence of $k_y$ on $q_{\rm s0}$ (\emph{upper panel}),
together with the corresponding growth rate $\lambda$ (\emph{lower panel}).
}\label{pk_vs_qs}\end{figure}

Earlier work has suggested that the eigenmodes of NEMPI can be
three-dimensional (BKR).
This could render two-dimensional calculations inadequate
if the first excited mode were indeed three-dimensional.
However, it turns out that the wavelength of the eigenmode in the
direction of the field increases as $\qs$ decreases.
In BKR, where three-dimensional ($y$-dependent) solutions to NEMPI
were first reported, $\qs$ was chosen to be around $10$,
and the fastest growing mode was indeed three-dimensional.
In \Fig{b003_ly2pi_qs0} we show that the effective wavenumber of the variation
of the field in the $y$ direction decreases with decreasing values of $\qs$.
This is shown quantitatively in \Fig{pk_vs_qs}, where we plot the dependence
of the typical value of the field-aligned wavenumber, $k_y$,
on the value of $\qsz$.
Here, $k_y$ is evaluated in a layer near the surface.

We find that the typical value of $k_y$ grows
with increasing values of $\qsz$.
In addition, we find that the growth rate of the instability, $\lambda$,
increases with $\qsz$ approximately linearly once $\qsz$
exceeds a value of around two.
The fact that ${k_y\to0}$ as ${\qsz\to0}$ is significant, because
BKKR and also K\"apyl\"a et al.\ (2012) found from
simulations that $\qsz\approx0$.
In that case, the characteristic length scale along
the direction of the field
becomes infinite and the calculation essentially two-dimensional.
Conversely, when studying NEMPI in two dimensions,
changing the value of $\qsz$ has no
effect on structure formation and the growth rate; see \Tab{2d3d}.
However, it is now clear that this is an artifact of restricting the
solutions to be two-dimensional.

\begin{table}[t!]\caption{
Comparison of normalized growth rates, $\lambda H_{\rho}^2/\etat$,
for different values of $\qs$,
for a three (3D) and two-dimensional (2D) simulation ($L_y\to\infty$).
}
\tabcolsep=12pt
{\begin{tabular}{c ccc}
\hline\noalign{\smallskip}
$\lambda H_\rho^2/\etat$ & $L_y/H_\rho$ & $\qs=0$ & $\qs=20$ \\[1.5pt]
\hline\noalign{\smallskip}
3D &   $2$    & 11 & 158 \\
2D & $\infty$ & 11 & 11 \\[1.5pt]
\hline
\label{2d3d}
\end{tabular}}
\end{table}

\begin{figure} %[b!]
\begin{center}
\includegraphics[width=\columnwidth]{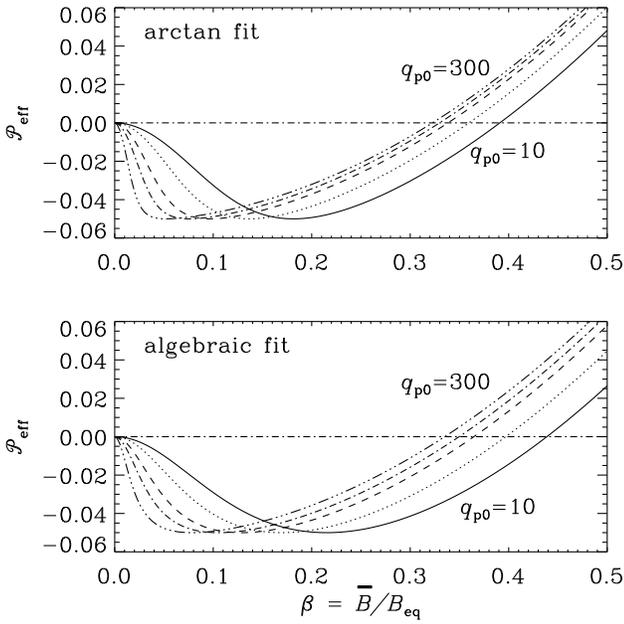}
\end{center}\caption[]{
\emph{Top}: comparison of original fit curves with different $\qpz$ values
(10, 20, 50, 100, 300) and a fixed minimum, showing similar
zero crossing values for a large $\qpz$ range.
\emph{Bottom}: same as above, but now using an algebraic fit, giving larger spacing between the curves.
}\label{degen}
\end{figure}

\subsection{Approximate degeneracy in the $\vec{q}_{\bf p}$ fit formula}

We mentioned in the introduction that recent attempts to determine
$\qpz$ from simulations faced the difficulty that the fit formula
\eq{qp} possesses an approximate degeneracy in that we can obtain
a similarly looking dependence of the effective mean magnetic pressure,
\EQ
\Peff(\beta)=\half[1-\qp(\beta)] \, \beta^2,
\EN
over a wide range of values
of $\qpz$ by adjusting the value of $\betap$ correspondingly.
This can be seen in \Fig{degen}, where we show $\Peff(\beta)$ using
either the arctan fit (upper panel) or the algebraic fit (lower panel)
for parameters that result in the same value of $\Pmin$ for a range
of values of $\qpz$.
Note that the position where $\Peff$ becomes positive, i.e.\ the critical
value defined by $\Peff(\betacrit)=0$, is rather similar in all cases.
This approximate degeneracy is particularly obvious for the arctan fit,
and less so for the algebraic fit.
However, in both cases the form of $\Peff$ near ${\beta\to0}$ changes
significantly.
Therefore, the approximate degeneracy would be lifted if one 
could determine $\qpz$ from the behavior of $\qp$ near $\beta=0$.
However, near $\beta=0$ the DNS have large errors.
It is therefore better to
measure the normalized minimum effective magnetic pressure,
$\Pmin=\half\min[(1-\qp)\beta^2]$, and its position, $\betamin$.
For the algebraic fit we then obtain the fit parameters
\begin{equation}
\betap=\betamin^2\left/\sqrt{-2\Pmin},\right.\;\;
\betastar=\betap+\sqrt{-2\Pmin},
\end{equation}
where we have introduced the parameter $\betastar^2=\qpz\betap^2$
in a modified representation
\begin{equation}
\qp(\beta)={\betastar^2\over\betap^2+\beta^2},
\label{qpbeta2b}
\end{equation}
which is preferable over \Eq{qpbeta2} in circumstances
where $\betastar^2=\qpz\betap^2$ is approximately constant.
This appears to be the case in recent DNS (BKKR, Kemel et al.\ 2012),
where $\betastar\approx0.2$ and $0.3$ in the absence and presence of
small-scale dynamo action, respectively.

\begin{figure}[t!]\begin{center}
\includegraphics[width=.95\columnwidth]{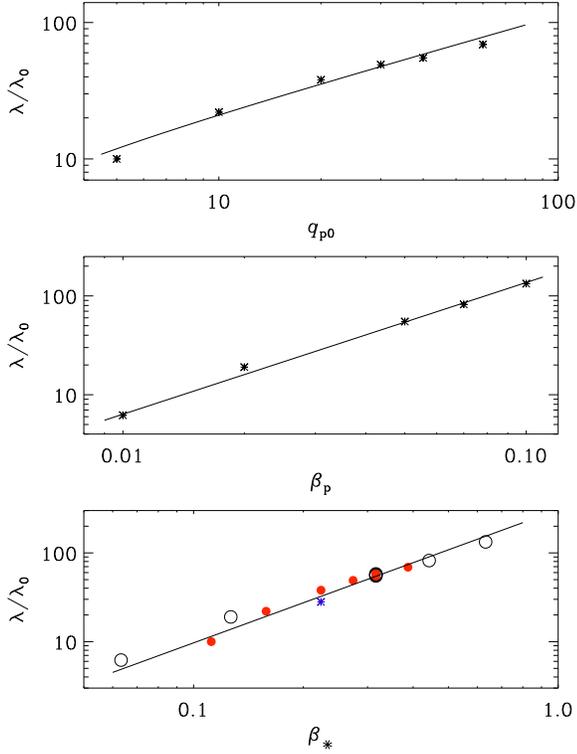}
\end{center}
\vskip-3mm
\caption[]{
Dependence of the normalized growth rate $\lambda/\lambda_0$,
with $\lambda_0=(\nut+\etat)/H_\rho^2$,
on $\qpz$ for $\betap=0.05$ (\emph{upper panel}), 
on $\betap$ for $\qpz=40$ (\emph{middle panel}),
and on $\betastar=\qpz^{1/2}\betap$ (\emph{lower panel})
for $\betap=0.05$ (red), $\qpz=40$ (circles),
and $\betap=0.1$, $\qpz=5$ (blue).
Solid lines represent approximate fits given by
$9(\qpz-1)^{0.7}$, $(\betap/0.0015)^{1.3}$, and
$(\betastar/0.02)^{3/2}$, respectively.
}\label{pBsBp}\end{figure}

We have computed mean-field models for different combinations of
parameters using the algebraic fit.
We find that the resulting growth rate $\lambda$ depends
on the functional form of $\Peff(\beta)$ near $\beta=\betamin$,
which manifests itself in a dependence on both $\qpz$ and $\betap$; see
\Fig{pBsBp}.
The lower panel of this figure suggests that the dependence of the growth rate 
on both parameters can be collapsed onto a single dependence on $\betastar$. 
This underlines the usefulness of \Eq{qpbeta2b} as a fit formula.
As argued above, this dependence is best constrained by the fit parameters
$\betamin$ and $\Pmin$.

\subsection{Onset of NEMPI}

With a given prescription of $\qp(\beta)$, assuming here $\qs=0$,
we can now compute two-dimensional mean-field models.
Our goal is to obtain a simple formula that can tell us
how large the growth rate of the instability is, and what the
critical condition for the onset of the instability is.
Not much is known about the linear stability properties of NEMPI,
so we have to rely on numerical determinations of the growth rates
for different wavelengths for different parameters to obtain an
approximate representation of the dispersion relation.
Earlier work of Kemel et al.\ (2011) has suggested a relation of the form
\EQ
\lambda=\Phi(g/\cs^2, \qpz, \betap, ...)
-\nut k_\nu^2-\etat k_\eta^2,
\label{fit1}
\EN
where $k_\nu$ and $k_\eta$ are effective wavenumbers quantifying
the effects of turbulent viscosity and turbulent magnetic diffusivity,
$\Phi$ is a function of
the inverse scale height, $H_\rho^{-1}=g/\cs^2$, and
other parameters describing the functional form of $\qp$.

\begin{figure}[t!]\begin{center}
\includegraphics[width=.95\columnwidth]{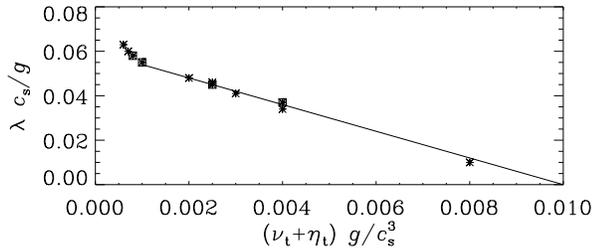}
\end{center}\caption[]{
Dependence of $\lambda$ on $\nut+\etat$, normalized appropriately
in terms of $g$ and $\cs$.
The negative slope gives $\kd^2$ with $\kd\approx2.5/H_\rho$,
where $H_\rho=\cs^2/g$.
}\label{lamfit2a}\end{figure}

We now need to determine the various unknowns.
We begin by determining $k_\nu$ and $k_\eta$ by varying
either only $\nut$ or only $\etat$ at a time.
It turns out that $k_\nu=k_\eta\equiv\kd$ suffices.
In this way we obtain a linear fit for the growth rate
${\lambda=\const-(\nut+\etat)\kd^2}$, giving us $\kd^2$ as the
slope of this graph; see \Fig{lamfit2a}.
We find $\kd\approx2.5/H_\rho$, superseding earlier
results by Kemel et al.\ (2011) for a different $\Beq(z)$ profile.

\begin{figure}[t!]\begin{center}
\includegraphics[width=.95\columnwidth]{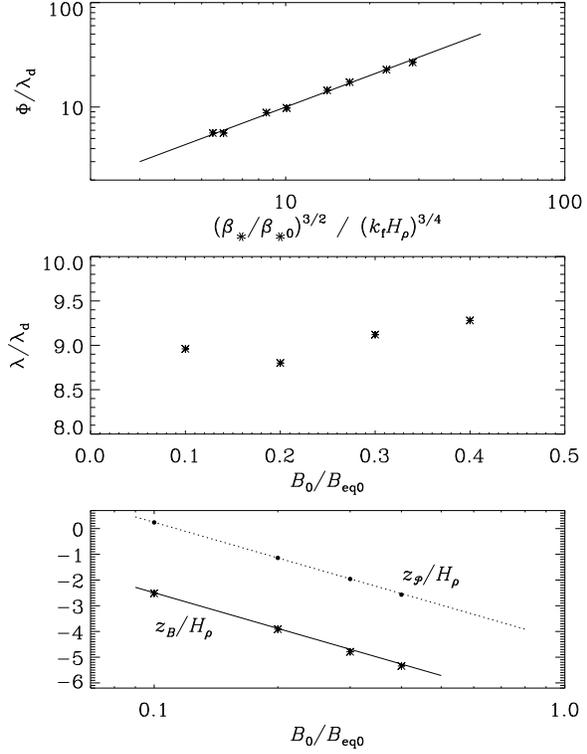}
\end{center}
\vskip-1mm
\caption[]{
Dependence of $\Phi=\lambda+(\nut+\etat)\kd^2$ on
$(\kf H_\rho)^{-3/4}$ (\emph{upper panel}),
$\lambda$ on $B_0/\Beqz$ (\emph{middle panel}),
as well as $z_B/H_\rho$ and $z_{\cal P}/H_\rho$ on $B_0/\Beqz$
(\emph{lower panel}).
Here, $\betastarz=0.0083$ is a fit parameter and
$\lambda_{\rm d}=(\nut+\etat)\kd^2$ is used for normalization.
%\vspace{-1.5mm}
}\label{lamfit2b}
\end{figure}

Accepting now the fit parameter $\kd$ as
measured, we can proceed to determining the dependence of
$\Phi$ on $H_\rho$ (top panel of \Fig{lamfit2b}).
A convenient non-dimensional quantity is $\kf H_\rho$, where $\kf$
is the wavenumber of the energy-carrying eddies of the turbulence,
which is related to $\etat=\urms/3\kf$ with $\urms=\Beqz/\rho_0^{1/2}$.
Note that $\Phi\propto(\kf H_\rho)^{-3/4}$.
Combining this with the $\betastar^{3/2}$ scaling of \Fig{pBsBp}, we suggest
\EQ
\lambda\approx\left[(\betastar/\betastarz)^{3/2}(\kf H_\rho)^{-3/4}-1\right]
(\nut+\etat)\kd^2,
\label{fit1}
\EN
with $\betastarz\approx0.008$ being yet another fit parameter.
Interestingly, $\lambda$ is independent of the imposed
field strength, $B_0/\Beqz$, provided the bulk of the eigenmode
($z=z_B$) fits well within the domain (middle panel of \Fig{lamfit2b}).
This has here been achieved by adjusting the positions of the
boundaries, $z_{\rm top}$ and $z_{\rm bot}$.
Indeed, as $B_0/\Beqz$ is increased, $z_B$ is found to decrease
approximately like $z_B/H_\rho\approx-2\ln B_0/\Beqz+\const$.
It turns out that $z_B$ is about 2--3 scale heights below the location
$z_{\cal P}$ where $\Peff(z)$ attains its minimum value.
Contrarily, when $z_{\cal P}>z_{\rm bot}>z_B$ or when
$z_{\rm top}<z_B$ NEMPI will depend on respectively
$\beta_{\rm bot}$ or $\beta_{\rm top}$,
when $z_{\cal P}<z_{\rm bot}$ there is no instability.

Some comments about the horizontal dimensions are in order.
In all cases with $\qsz=0$, we find that in three-dimensional
calculations with finite $y$ extent, the value of $L_y$
does not affect the growth rates.
On the other hand, doubling the $x$ extent yields two pairs
of rolls.
This has also been confirmed for DNS; see Kemel et al.\ (2012).

\section{Conclusions}

The present work has clarified a number of puzzling aspects of NEMPI.
Firstly, it is now clear that we can proceed with two-dimensional
mean-field simulations as long as we know that $\qsz=0$ (or negative).
However, this may not always be the case.
The fact that three-dimensional structures can emerge from NEMPI
was initially thought to be an interesting aspect, because it could
readily explain the formation of bipolar regions (BKR).
However, given that simulations now indicate that $\qs\approx0$ (or
perhaps even negative), this proposal would thus not be an option,
unless some other as yet unexplored effect begins to play a role.
In principle, all turbulent transport processes are nonlocal and
must be described by a convolution with the mean field rather
than a multiplication (Brandenburg et al.\ 2008).
In Fourier space, the convolution corresponds to a multiplication
with a wavenumber-dependent turbulent transport coefficient.
Thus, the idea of explaining bipolar regions would again
become viable if this effect only existed at small and
intermediate length scales.
Clarifying this would be a task for future simulations, because none of the
currently available techniques are yet equipped to address this possibility.

Next, we have seen that the degeneracy in the fit formula used for
$\qp(\beta)$ and $\Peff(\beta)$ is significant in that
different combinations of $\qpz$ and $\betap$ result in similar
values of $\min(\Peff)$ and $\betacrit$, but the growth
rates can still be quite different.
This means that it is not sufficient to measure only
$\min(\Peff)$ and $\betacrit$.
Instead, to characterize the functional form of $\Peff(\beta)$
more accurately, we need some other characteristics to represent the
dependence of this function near $\beta=0$.
One such possibility is to use the field strength $\betamin$ for which
the minimum of the effective magnetic pressure is reached.

Knowing the value of $\betamin$ has particular relevance in determining
the height where NEMPI occurs.
For a given value of the imposed field strength $B_0$, the condition
$B_0/\Beq(z_{\cal P})=\betamin$ determines the height $z_{\cal P}$, where
the effective magnetic pressure attains a minimum, and thus the height
$z_B$, which tends to be 2--3 scale heights below $z_{\min}$;
see the lower panel of \Fig{lamfit2b}.
Therefore, the value of $B_0$ does not directly affect the growth rate
of NEMPI.

Finally, we have tried to establish an approximate dispersion relation to estimate the growth rate of NEMPI as a function of turbulent viscosity,
turbulent magnetic diffusivity, mean field strength, and the strength
of stratification.
This formula may serve as a first orientation and can hopefully be
improved further with future simulations.
This formula can also be useful in connection with analytic estimates
concerning the regimes when NEMPI is expected in DNS or under other more
realistic circumstances.

\acknowledgements
We thank the anonymous reviewer for making useful suggestions.
We acknowledge the NORDITA dynamo programs of 2009 and 2011 for
providing a stimulating scientific atmosphere.
Computing resources provided by the Swedish National Allocations Committee
at the Center for Parallel Computers at the Royal Institute of Technology in
Stockholm, the National Supercomputing Center in Link\"oping,
and the High Performance Computing Center North in Ume{\aa}.
This work was supported in part by the European Research Council under the AstroDyn project 227952 and the Swedish Research Council grant 621-2011-5076.

\end{document}